\documentclass[10pt,conference, hidelinks]{IEEEtran}
\IEEEoverridecommandlockouts

\usepackage{cite}
\usepackage{amsmath,amssymb,amsfonts}
\usepackage{graphicx}
\usepackage{float}                      
\usepackage{textcomp}
\usepackage[dvipsnames]{xcolor}
\usepackage[acronym,shortcuts]{glossaries}
\usepackage{bm}
\usepackage{caption}
\usepackage{subcaption}
\usepackage{relsize}
\usepackage{soul}
\usepackage{algorithm,algpseudocode}
\usepackage{outlines}
\usepackage{lipsum,comment}
\usepackage{mathtools}
\usepackage{tabularx,flushend}
\usepackage{amsthm}

\theoremstyle{definition}

\def\BibTeX{{\rm B\kern-.05em{\sc i\kern-.025em b}\kern-.08em
T\kern-.1667em\lower.7ex\hbox{E}\kern-.125emX}}

\newcommand{\trans}[0]{^{\mathsf{T}}}

\newcommand{\herm}[0]{^{\mathsf{H}}}

\newacronym{RPE}{RPE}{radar parameter estimation}
\newacronym{OTFS}{OTFS}{orthogonal time frequency space}
\newacronym{AFDM}{AFDM}{affine frequency division multiplexing}
\newacronym{MIMO}{MIMO}{multiple-input multiple-output}
\newacronym{SISO}{SISO}{single-input single-output}
\newacronym{ISAC}{ISAC}{integrated sensing and communications}
\newacronym{DISAC}{DISAC}{distributed ISAC}
\newacronym{3D}{3D}{three-dimensional}
\newacronym{2D}{2D}{two-dimensional}
\newacronym{1D}{1D}{one-dimensional}
\newacronym{RX}{RX}{receiver}
\newacronym{TX}{TX}{transmitter}
\newacronym{BF}{BF}{beamforming}
\newacronym{mmWave}{mmWave}{millimeter-wave}
\newacronym{SotA}{SotA}{state-of-the-art}
\newacronym{ULA}{ULA}{uniform linear array}
\newacronym{QAM}{QAM}{quadrature amplitude modulation}
\newacronym{ISFFT}{ISFFT}{inverse symplectic finite Fourier transform}
\newacronym{SFFT}{SFFT}{symplectic finite Fourier transform}
\newacronym{AWGN}{AWGN}{additive white Gaussian noise}
\newacronym{OFDM}{OFDM}{orthogonal frequency division multiplexing}
\newacronym{OCDM}{OCDM}{orthogonal chirp division multiplexing}
\newacronym{BS}{BS}{base station}
\newacronym{UE}{UE}{user equipment}
\newacronym{DFT}{DFT}{discrete Fourier transform}
\newacronym{IDFT}{IDFT}{inverse discrete Fourier transform}
\newacronym{TD}{TD}{time-domain}
\newacronym{wlg}{wlg}{without loss of generality}
\newacronym{CP}{CP}{cyclic prefix}
\newacronym{DAFT}{DAFT}{discrete affine Fourier transform}
\newacronym{IDAFT}{IDAFT}{inverse discrete affine Fourier transform}
\newacronym{CPP}{CPP}{\textit{chirp-periodic} prefix}
\newacronym{IDZT}{IDZT}{inverse discrete Zak transform}
\newacronym{DZT}{DZT}{discrete Zak transform}
\newacronym{ICI}{ICI}{inter-carrier interference}
\newacronym{BER}{BER}{bit error rate}
\newacronym{DoF}{DoF}{degrees-of-freedom}
\newacronym{FD}{FD}{full-duplex}
\newacronym{SIMO}{SIMO}{single-input multiple-output}
\newacronym{MISO}{MISO}{multiple-input single-output}
\newacronym{AoD}{AoD}{angle-of-departure}
\newacronym{AoA}{AoA}{angle-of-arrival}
\newacronym{RF}{RF}{radio frequency}
\newacronym{SIM}{SIM}{stacked intelligent metasurfaces}
\newacronym{FIM}{FIM}{Fisher information matrix}
\newacronym{FPGA}{FPGA}{field programmable gate array}
\newacronym{UPA}{UPA}{uniform planar array}
\newacronym{CC}{CC}{communication-centric}
\newacronym{I/O}{I/O}{input-output}
\newacronym{iid}{i.i.d.}{independent and identically distributed}
\newacronym{IoT}{IoT}{internet of things}
\newacronym{V2X}{V2X}{vehicle-to-everything}
\newacronym{NTN}{NTN}{non-terrestrial network}
\newacronym{LEO}{LEO}{low-earth orbit}
\newacronym{THz}{THz}{terahertz}
\newacronym{EM}{EM}{electromagnetic}
\newacronym{STAR-RIS}{STAR-RIS}{simultaneously transmitting and reflecting reconfigurable intelligent surface}
\newacronym{DoA}{DoA}{direction-of-arrival}
\newacronym{DD}{DD}{doubly-dispersive}
\newacronym{ODDM}{ODDM}{orthogonal delay-Doppler division multiplexing}
\newacronym{LoS}{LoS}{line-of-sight}
\newacronym{NLoS}{NLoS}{non-line-of-sight}
\newacronym{6G}{6G}{sixth generation}
\newacronym{MPDD}{MPDD}{metasurfaces-parameterized DD}
\newacronym{GaBP}{GaBP}{Gaussian belief propagation}
\newacronym{MSE}{MSE}{mean-squared-error}
\newacronym{sIC}{soft IC}{soft interference cancellation}
\newacronym{soft RG}{soft RG}{soft replica generation}
\newacronym{BG}{BG}{belief generation}
\newacronym{SGA}{SGA}{scalar Gaussian approximation}
\newacronym{CLT}{CLT}{central limit theorem}
\newacronym{PDF}{PDF}{probability density function}
\newacronym{QPSK}{QPSK}{quadrature phase-shift keying}
\newacronym{LMMSE}{LMMSE}{linear minimum mean square error}
\newacronym{SNR}{SNR}{signal-to-noise ratio}
\newacronym{QoS}{QoS}{quality of service}
\newacronym{CoV}{CoV}{calculus of variations}
\newacronym{CAPA}{CAPA}{continuous aperture array}
\newacronym{FCAPA}{FCAPA}{flexible continuous aperture array}
\newacronym{GL}{GL}{Gauss-Legendre}
\newacronym{DDC MIMO}{DDC MIMO}{DD continuous MIMO}
\newacronym{B5G}{B5G}{beyond fifth generation}
\newacronym{VR}{VR}{virtual reality}
\newacronym{XR}{XR}{extended reality}
\newacronym{ITN}{ITN}{intelligent traffic networks}
\newacronym{SAGIN}{SAGIN}{space-air-ground integrated network}
\newacronym{UAV}{UAV}{unmanned aerial vehicle}
\newacronym{MUSIC}{MUSIC}{multiple signal classification}
\newacronym{ICC}{ICC}{integrated communication and computing}
\newacronym{SINR}{SINR}{signal-to-interference-plus-noise ratio}
\newacronym{WSR}{WSR}{weighted sum rate}
\newacronym{ARPU}{ARPU}{average rate per user}
\newacronym{BCD}{BCD}{block coordinate descent}
\newacronym{PDE}{PDE}{partial differential equation}
\newacronym{EL}{EL}{Euler-Lagrange}
\newacronym{TCA}{TCA}{tightly coupled array}
\newacronym{ELAA}{ELAA}{extremely large-aperture arrays}
\newacronym{LIS}{LIS}{large intelligent surface}
\newacronym{CSI}{CSI}{channel state information}
\newacronym{RIS}{RIS}{reconfigurable intelligent surface}
\newacronym{KKT}{KKT}{Karush-Kuhn-Tucker}
\newacronym{TO}{TO}{time offset}
\newacronym{CFO}{CFO}{carrier frequency offset}
\newacronym{BGaBP}{BGaBP}{bivariate Gaussian belief propagation}
\newacronym{RRC}{RRC}{root raised cosine}
\newacronym{TDD}{TDD}{time-division duplex}
\newacronym{ZF}{ZF}{zero-forcing}
\newacronym{LS}{LS}{least squares}
\newacronym{MF}{MF}{matched filter}
\newacronym{IC}{IC}{interference cancellation}
\newacronym{RMSE}{RMSE}{root mean-squared error}
\newacronym{CRLB}{CRLB}{Cram\'er Rao lower bound}
\newacronym{JCDE}{JCDE}{joint channel and data estimation}

\begin{document}

\title{Joint Synchronization and Radar Parameter Estimation for {\color{black} OFDM-based  DISAC} Systems}

\author{\IEEEauthorblockN{Niclas~F\"uhrling$^\dag$, Hyeon~Seok~Rou$^\dag$, Kuranage~Roche~Rayan~Ranasinghe$^\dag$, \\ Giuseppe~Thadeu~Freitas~de~Abreu$^\dag$ and Nuria~Gonz{\'a}lez-Prelcic$^\ddag$}\\[-2ex]
\IEEEauthorblockA{\textit{$^\dag$School of Computer Science and Engineering, Constructor University, Bremen, Germany} \\ \textit{$^\ddag$Department of Electrical and Computer Engineering, University of California San Diego, La Jolla, CA, USA} \\[0.5ex]
\{nfuehrling, hrou, kranasinghe, gabreu\}@constructor.university, ngprelcic@ucsd.edu}
\vspace{-4ex}
}

\maketitle

\begin{abstract}
We {\color{black} propose} a novel approach to the synchronization paradigm in \ac{DISAC} systems in \ac{DD} channel environments via a joint synchronization and radar parameter estimation framework.
The proposed method exploits the structure of the system model, which can be linearized in order to apply a  bivariate \ac{GaBP} algorithm that {\color{black} jointly} estimates the \ac{TO} and \ac{CFO} of each \ac{BS}, as well as the delay and Doppler parameters of the \ac{DD} channel in conventional \ac{OFDM} systems.
Simulation results demonstrate the effectiveness of the proposed algorithm, {\color{black} showing that the radar parameter estimates ($i.e.$, range and velocity) and synchronization parameter estimates ($i.e.$, \ac{TO} and \ac{CFO}) approach the \ac{CRLB} at moderate-to-high \ac{SNR} regimes.}
\end{abstract}

\begin{IEEEkeywords}
Distributed ISAC (DISAC), clock synchronization, radar parameter estimation, Gaussian belief propagation
\end{IEEEkeywords}

\glsresetall

\vspace{-1ex}
\section{Introduction}
\vspace{-0.5ex}

\Ac{ISAC} is a key technology for future wireless networks, enabling the simultaneous use of communication and sensing functionalities over shared resources, which can be seen as one of the main applications in \ac{B5G} and \ac{6G} systems \cite{Wild_2021,Qaiser_2026,Zhang_2026,Lu_2024}.
Recently, in contrast to conventional mono-static \ac{ISAC} systems, where a single node performs both sensing and communication tasks, there has been a growing interest in \ac{DISAC} architectures, where multiple spatially distributed nodes collaborate to achieve enhanced performance \cite{Guo_2025,Strinati_2025,Meng_2024}. 
However, the distributed nature of \ac{DISAC} systems introduces new challenges, since each node operates with an independent clock, {\color{black} leading to unknown \ac{TO} and \ac{CFO} between the nodes.
Without proper synchronization, the estimation becomes infeasible, since the synchronization parameters are absorbed into the effective channel parameters and thus cannot be separately identified from them} \cite{Brunner_2024,Han_2025_IEEE,han_2025}.
Additionally, while \ac{OFDM} is a widely adopted modulation scheme in \ac{ISAC} systems, the fact that high mobility and large bandwidths in \ac{6G} systems give rise to \ac{DD} fading channels \cite{Rou_2024}, which are characterized by both delay and Doppler spreads, further complicates the synchronization and channel estimation process{\color{black}, which has been less addressed in the \ac{SotA} \ac{DISAC} literature.}

Conventional \ac{DISAC} solutions typically assume perfect synchronization or rely on synchronization treated as a separate preliminary step.
To name a few examples, \cite{han_2025} proposed a {\color{black} time-frequency} synchronization in \ac{OFDM} \ac{DISAC} systems, where an {\color{black} over-the-air} framework is designed to perform the \ac{TO} and \ac{CFO} estimation, {\color{black} without jointly recovering the delay-Doppler channel parameters.}
An example of \ac{JCDE} was proposed in \cite{RanasingheTWC2025JCDE}, where a \ac{GaBP} was designed to perform \ac{JCDE} jointly in a conventional \ac{ISAC} system, without taking into account the effect of synchronization.

In light of the above, since, to the best of our knowledge, no existing work has addressed the problem of joint synchronization and channel estimation in \ac{DISAC} systems operating in \ac{DD} fading environments, we propose a novel method for the joint estimation of the synchronization and {\color{black} delay-Doppler} parameters in \ac{OFDM}-based \ac{DISAC} systems.
Specifically, assuming known path gains and a coarse initialization of the delay-Doppler parameters, we focus on the iterative refinement of the synchronization and delay-Doppler estimates via a bivariate GaBP.
 
The rest of this paper is organized as follows.
First, the system model is offered in Section \ref{sec:system_model}.
Then, in Section \ref{sec:proposed}, the proposed method for the joint estimation of the synchronization parameters and the delay Doppler parameters is introduced.
Finally, a comparison of the proposed \ac{GaBP}-based scheme with the \ac{CRLB} is presented in Section \ref{sec:results}.


\section{System Model}
\label{sec:system_model}

\begin{figure}[H]
    \centering
    \includegraphics[width=\columnwidth]{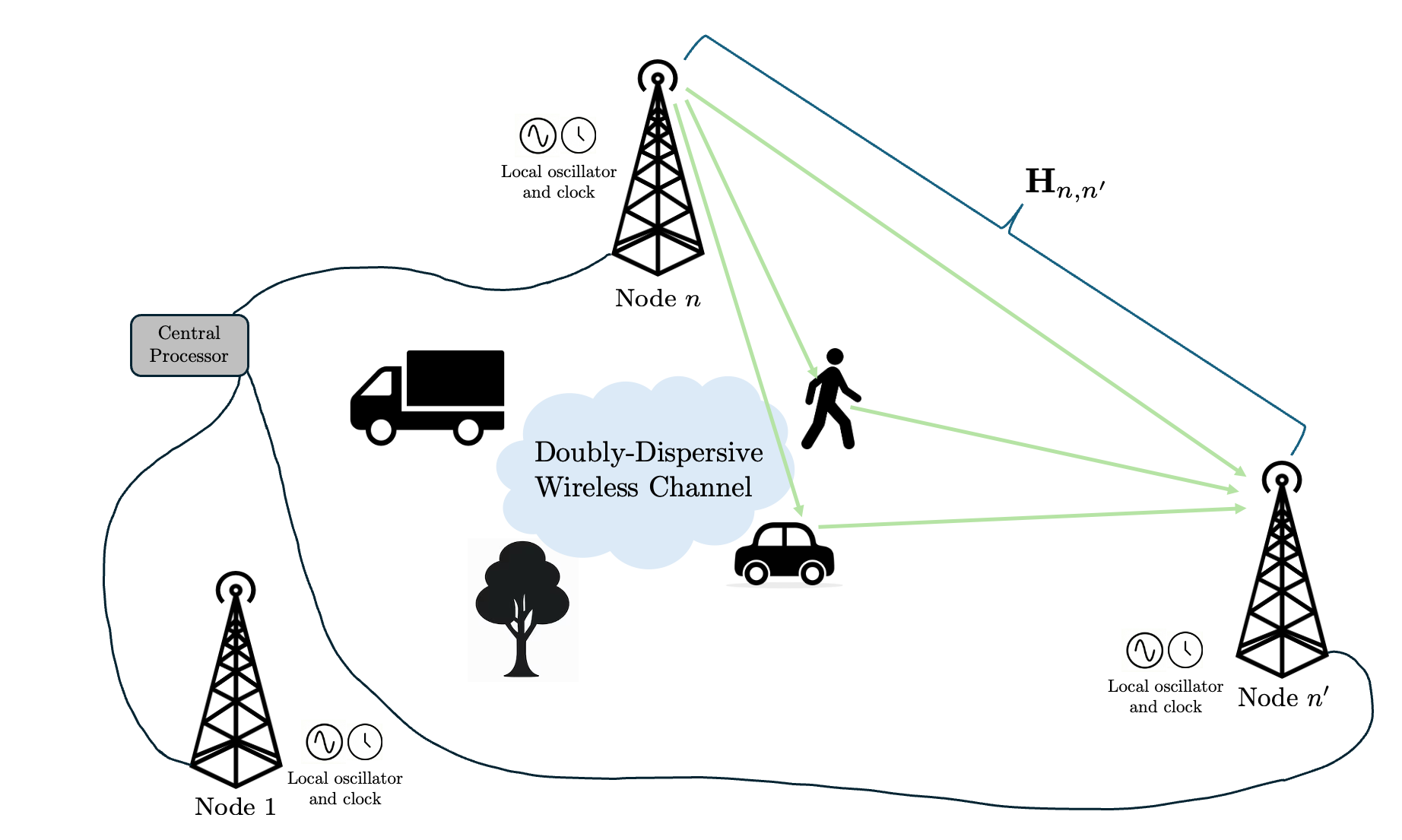}
    \caption{Illustration of a \ac{DISAC} system with $N$ distinct single-antenna nodes.}
    \label{fig:system_model}
\end{figure}

Consider a \ac{DISAC} network comprising $N$ single-antenna transmit nodes (or equivalently, \acp{BS}) as commonly assumed in the \ac{SotA} \cite{han_2025,Guo_2025,Meng_2024} and illustrated in Figure \ref{fig:system_model}. 
Unlike what is assumed in most conventional \ac{SotA} methods, each \ac{BS} is subject to a \ac{TO} and \ac{CFO}, which requires the nodes to perform synchronization before communication and other applications can be performed.
Instead of treating synchronization as a separate preliminary step, the nodes can therefore perform synchronization and channel estimation jointly from the same received signals, as described by the following signal model.

\newpage

\subsection{Transmit Signal Model}

Let the $k$-th data symbol transmitted by the $n$-th \ac{DISAC} node be denoted by $x_{k,n} \in \mathcal{X}$ for $k \in \{0,\ldots,K-1\}$ and $n \in \{1,\ldots,N\}$,  where $\mathcal{X} \subset \mathbb{C}$ is the complex constellation set, with cardinality $|\mathcal{X}| = D$, $i.e.$, $D$-QAM or $D$-PSK.
Then, the $K \times 1$ data symbol vector to be transmitted from the $n$-th node can be defined as $\mathbf{x}_n \triangleq [x_{0,n}, \ldots, x_{K-1,n}]\trans \in \mathbb{C}^{K \times 1}$.

Next, the baseband equivalent \ac{OFDM} transmit signal block with $K$ samples in the time domain is generated at each $n$-th node as
\begin{equation}
    \mathbf{s}_n = \mathbf{F}_K\herm \mathbf{x}_n \in \mathbb{C}^{K \times 1},
\end{equation}
where $\mathbf{s}_n \in \mathbb{C}^{K \times 1}$ is the transmit signal vector in the time domain and $\mathbf{F}_K \in \mathbb{C}^{K \times K}$ is the normalized $K$-point \ac{DFT} modulation matrix.

\subsection{Channel Model}

Consider a baseband equivalent point-to-point wireless channel with $P$ significant (resolvable) scattering propagation paths, characterized by its continuous-time transfer function expressed as
\begin{equation}
h(t,\tau) = \sum_{p=1}^{P} h_p \, e^{j 2 \pi \nu_p t} \, \delta(\tau - \tau_p),
\label{eq:cont_time_SISO_tvirf}
\end{equation}
where $h_p \in \mathbb{C}$ denotes the complex gain, $\tau_p \in \mathbb{R}$ denotes the propagation delay, and $\nu_p \in \mathbb{R}$ denotes the associated Doppler shift of the $p$-th path, and the combined influence of the path delays $\tau_p$ (frequency selectivity) and Doppler shifts $\nu_p$ (time selectivity) gives rise to the so-called \ac{DD} fading environment.
The propagation delay of path $p$ on any given link is determined by the corresponding path length $d_p \in [d_{\min}, d_{\max}]$ as
\begin{equation}
    \tau_p = \frac{d_p}{c} \in \mathbb{R}^+, \label{eq:tau_from_d}
\end{equation}
and the Doppler shift of path $p$ is determined by the radial velocity $v_p \in [-v_{\max}, v_{\max}]$ of the associated scatterer as
\begin{equation}
    \nu_p = \frac{v_p\, f_c}{c} \in \mathbb{R}. \label{eq:nu_from_v}
\end{equation}

The maximum delay spread and maximum Doppler spread of the channel are defined as $\tau_{\max} \in \mathbb{R}^+$ and $\nu_{\max} \in \mathbb{R}^+$, respectively, such that $\tau_p \in [0, \tau_{\max}]$ and $\nu_p \in [-\nu_{\max}, \nu_{\max}]$ for $p \in \{1,\ldots,P\}$, and the channel is considered underspread if $\tau_{\max} \nu_{\max} <\!\!< 1$, and overspread otherwise.

For convenience, let us also define the normalized delay indices and the normalized digital Doppler shift indices as
\begin{subequations}
    \begin{equation}
    \ell_p \triangleq \frac{\tau_p}{T_s},
\label{eq:normalized_delay} 
\end{equation}
\begin{equation}
    f_p \triangleq K \nu_p T_s = K\frac{\nu_p}{f_S},
\label{eq:normalized_doppler} 
\end{equation}
\label{eq:normalized_DD}
\end{subequations}
where $K$ is the total number of samples in the sequence, $T_s$ is the sampling period and $f_S = 1/T_s$ is the sampling frequency.

For a general case where both the integer and fractional normalized delay and Doppler shifts are considered, it was shown in \cite{rou2026afdmevolvingofdm6g} that an overall convolution matrix $\mathbf{H} = \sum_{p=1}^{P} h_p \mathbf{H}_p \in \mathbb{C}^{K \times K}$ that represents the \ac{DD} channel can be constructed as
\begin{equation}
\mathbf{H}_p \triangleq \mathbf{V}^{f_p} \cdot \mathbf{\Psi}(\ell_p) \in \mathbb{C}^{K \times K}.
\label{eq:Hp_FDFD_generalized}
\end{equation}

{\color{black}The effect of the normalized digital Doppler shifts $f_p$ based on the $K$-th roots-of-unity matrix is therefore modeled by the diagonal Doppler shift matrix $\mathbf{V}^{f_p}$, given by}
\begin{equation}
\mathbf{V} \triangleq \mathrm{diag}\Big(\Big[1, e^{-j2\pi\frac{1}{K}}, \cdots, e^{-j2\pi\frac{K-1}{K}}\Big]\Big) \in \mathbb{C}^{K \times K}, \label{eq:rootofunity_matrix}
\end{equation}
which when raised to the $f_p$-th power as $\mathbf{V}^{f_p}$, is the diagonal matrix representing the phase rotation due to the Doppler shift.

In addition, the fractional delay matrix $\mathbf{\Psi}(\ell_p)\in \mathbb{C}^{K \times K}$ captures the effect of the (possibly fractional) delay $\ell_p$ via pulse-shaped interpolation, defined element-wise as
\begin{equation}
[\mathbf{\Psi}(\ell_p)]_{i,j} = g(i - j - \ell_p),
\label{eq:fracdel_interp_matrix}
\end{equation}
where $g(\cdot)$ is the pulse-shaping filter impulse response, and $\mathbf{\Psi}(\ell_p)$ is generally dense and Toeplitz \cite{Proakis_2001}.

According to \cite{rou2026afdmevolvingofdm6g}, {\color{black} \eqref{eq:Hp_FDFD_generalized}} is the \ac{OFDM} specialization of the unified \ac{DD} channel model, for which the cyclic prefix guards against inter-block interference, while the cyclic wrap-around is negligible for the well-localized pulse considered here.
In the present work, a \ac{RRC} pulse shape with roll-off factor $\alpha = 0.3$ is employed.
Finally, let $\mathbf{H}_{n,n'} {\color{black} \in} \mathbb{C}^{K \times K}$ denote the \ac{DD} channel matrix at a node $n$ resulting from a transmission by a node $n'$ and reflected by $P_{n,n'}$ scattering propagation paths.

In a \ac{DISAC} architecture, {\color{black}following the \ac{SotA} \cite{han_2025},} each node $n$ operates with an independent clock characterized by an unknown timing offset $\epsilon_{t,n} \in \mathbb{R}$ (seconds) and carrier frequency offset $\epsilon_{f,n} \in \mathbb{R}$ (Hz). 
Since no global synchronization is assumed, only \emph{relative} offsets between transmitter $n'$ and receiver $n$
affect the received signal.
We assume that $\epsilon_{t,n}$ and $\epsilon_{f,n}$ are constant over one $K$-sample block.

We define the relative timing and frequency offsets as
\begin{align}
\Delta\epsilon_{t,n,n'} &= \epsilon_{t,n'} - \epsilon_{t,n}, \\
\Delta\epsilon_{f,n,n'} &= \epsilon_{f,n'} - \epsilon_{f,n}.
\end{align}

These relative offsets will be absorbed into the effective \ac{DD} channel parameters.

Finally, with $P_{n,n'}$ propagation paths between transmitter $n'$ and receiver $n$, the physical delay and Doppler of path $p$ can be expressed as $\tau_{p,n,n'}$ and $\nu_{p,n,n'}$, respectively.

In the presence of relative synchronization offsets, the effective delay and Doppler become
\begin{align}
\tau_{p,n,n'}^{\mathrm{eff}}
&=
\tau_{p,n,n'}
+
\Delta\epsilon_{t,n,n'}, \label{eq:tau_eff}\\
\nu_{p,n,n'}^{\mathrm{eff}}
&=
\nu_{p,n,n'}
+
\Delta\epsilon_{f,n,n'}.\label{eq:nu_eff}
\end{align}

Then, the corresponding normalized discrete delay and Doppler indices are
\begin{align}
\ell_{p,n,n'}^{\mathrm{eff}}
&= \frac{\tau_{p,n,n'}^{\mathrm{eff}}}{T_s}, \\
f_{p,n,n'}^{\mathrm{eff}}
&= K \nu_{p,n,n'}^{\mathrm{eff}} T_s.
\end{align}

Next, the expression that follows generalizes the point-to-point \ac{DD} channel in \eqref{eq:Hp_FDFD_generalized} to the distributed unsynchronized \ac{ISAC} setting by
absorbing relative synchronization offsets into the path-dependent delay and Doppler parameters.

The discrete \ac{DD} channel operator\footnote{\color{black}Note that, for the sake of simplicity, the phase offset of a conventional bistatic channel \cite{Wu_2024} is neglected, but will be considered in a follow-up work.} is therefore
\begin{equation}
\mathbf{H}_{n,n'}
=
\sum_{p=1}^{P_{n,n'}}
h_{p,n,n'}
\mathbf{V}^{\,f_{p,n,n'}^{\mathrm{eff}}}
\mathbf{\Psi}\!\left(\ell_{p,n,n'}^{\mathrm{eff}}\right),
\label{eq:DISAC_DD_channel}
\end{equation}
where $\ell_{p,n,n'}^{\mathrm{eff}}$ denotes the discretized delay index corresponding to $\tau_{p,n,n'}^{\mathrm{eff}}$ and $f_{p,n,n'}^{\mathrm{eff}}$ denotes the normalized Doppler index corresponding to $\nu_{p,n,n'}^{\mathrm{eff}}$.



\subsection{Receive Signal Model}

Since all \ac{DISAC} nodes operate within the same time-frequency resources, the received signal at each \ac{ISAC} node is the summation of all the reflected signals from a given set of $P_{n,n'}$ scattering propagation
paths.

Therefore, leveraging the \ac{DD} channel matrix representation in \eqref{eq:DISAC_DD_channel}, the received baseband signal at a given $n$-th \ac{ISAC} node can be expressed as
\begin{align}
    \label{eq:n_th_nodeISAC}
    \mathbf{r}_n &= \sum_{n'=1}^N \mathbf{H}_{n,n'} \mathbf{s}_{n'} + \mathbf{w}_n  \in \mathbb{C}^{K \times 1} \nonumber \\
    & =\sum_{n'=1}^N \mathbf{H}_{n,n'} {\color{black} \mathbf{F}_K\herm} \mathbf{x}_{n'} + \mathbf{w}_n.
\end{align}

Next, demodulating the \ac{TD} baseband signal yields
\begin{align}
    \label{eq:I/O_rel_TD}
    \mathbf{y}_n &= {\color{black} \mathbf{F}_K} \mathbf{r}_{n}  \in \mathbb{C}^{K \times 1} \nonumber \\
    & = \sum_{n'=1}^N {\color{black} \mathbf{F}_K} \mathbf{H}_{n,n'} {\color{black} \mathbf{F}_K\herm} \mathbf{x}_{n'} + {\color{black} \mathbf{F}_K} \mathbf{w}_n.
\end{align}

Finally, by defining an equivalent noise vector $\tilde{\mathbf{w}}_n \triangleq {\color{black} \mathbf{F}_K} \mathbf{w}_n$ with the same statistics as $\mathbf{w}_n$ and an equivalent channel matrix $\bar{\mathbf{H}}_{n,n'} \triangleq {\color{black} \mathbf{F}_K} \mathbf{H}_{n,n'} {\color{black} \mathbf{F}_K\herm}$, the end-to-end input-output relationship can be expressed as
\begin{equation}
    \label{eq:I/O_rel_final}
    \mathbf{y}_n = \sum_{n'=1}^N \bar{\mathbf{H}}_{n,n'} \mathbf{x}_{n'} + \tilde{\mathbf{w}}_n \in \mathbb{C}^{K \times 1},
\end{equation}
which allows for the processing of the signals at each individual $n$-th \ac{ISAC} node.


\section{Proposed Joint Estimation Framework}
\label{sec:proposed}

The proposed framework targets the joint estimation of two groups of unknowns from the received signals $\{\mathbf{y}_n\}_{n=1}^N$, namely the physical delay-Doppler parameters $\{\tau_{p,e}, \nu_{p,e}\}$ and the synchronization parameters $\{\epsilon_{t,n}, \epsilon_{f,n}\}$.
Since the \ac{GaBP} solver operates on a linearized observation model, it requires an initial expansion point in the form of a coarse estimate, which is then iteratively refined through re-linearization.
For simplicity, this work focuses on the iterative refinement, whereas the coarse initialization and the data-driven estimation of the path gains are deferred to a follow-up work and are therefore not exercised here.

Furthermore, following the \ac{SotA} \cite{han_2025}, the framework operates under the following assumptions:
\begin{enumerate}
    \item \textbf{Pilot-aided transmission.} The transmitted symbol vectors $\mathbf{x}_{n'}$ are known to the central processor for all $n'$.
    \item \textbf{Known path gains and emulated coarse initialization.} The complex path gains $\{h_{p,e}\}$ are assumed available, and the coarse physical initialization is modeled as $\hat{\tau}_{p,e}^{(0)} = \tau_{p,e} + n_\tau$ and $\hat{\nu}_{p,e}^{(0)} = \nu_{p,e} + n_\nu$, with $n_\tau$ and $n_\nu$ zero-mean Gaussian, rather than produced via an additional estimation stage.
\end{enumerate}
Under these assumptions, the contribution of this paper is the bivariate \ac{GaBP} refinement, whose performance against the corresponding bounds is evaluated in Section~\ref{sec:results}.

\subsection{Network Topology and Reference Frame}
\label{sec:topology}

The network consists of $N$ single-antenna \ac{ISAC} nodes. 
It can be assumed that all $\binom{N}{2}$ distinct node pairs can exchange signals, yielding the set of undirected physical edges
\begin{subequations}
\begin{equation}
        \mathcal{E} = \bigl\{\{n,n'\} : 1 \le n < n' \le N\bigr\}, 
\end{equation}
    \begin{equation}
         |\mathcal{E}| = \frac{N(N-1)}{2}.
    \end{equation}
\end{subequations}

Each undirected edge $e \in \{1,\ldots,|\mathcal{E}|\}$ is expanded into two directed links, yielding a set $\mathcal{L}$ of $|\mathcal{L}| = N(N-1)$ directed links. 
For a directed link $l \in \mathcal{L}$, we denote the receiver by $n_l$, the transmitter by $n'_l$, and the index of the corresponding undirected edge by $e_l$.

For the sake of global synchronization, node $1$ is designated as the absolute reference clock such that
\begin{equation}
    \epsilon_{t,1} = 0, \quad \epsilon_{f,1} = 0.
    \label{eq:reference_node}
\end{equation}

The absolute timing and frequency offsets of all remaining $N-1$ nodes are unknown and can be collected in a global synchronization parameter vector, given by
\begin{equation}
    \boldsymbol{\theta}_{\mathrm{sync}} \triangleq
    \bigl[\epsilon_{t,2},\, \epsilon_{f,2},\, \ldots,\, \epsilon_{t,N},\, \epsilon_{f,N}\bigr]\trans
    \in \mathbb{R}^{2(N-1) \times 1}.
    \label{eq:theta_sync}
\end{equation}
The column index within $\boldsymbol{\theta}_{\mathrm{sync}}$ corresponding to the timing offset of node $n \geq 2$ is $c_n = 2(n-2)+1$, and for the frequency offset it is $c_n + 1$.

Furthermore, by \ac{TDD} channel reciprocity, the physical delay $\tau_{p,e}$ and Doppler $\nu_{p,e}$ of path $p$ on edge $e$ are identical for both transmission directions \cite{Liu_2020} and can be collected in a global physical parameter vector, given by
\begin{equation}
    \boldsymbol{\theta}_{\mathrm{phys}} \triangleq
    \bigl[\tau_{1,1},\, \nu_{1,1},\, \ldots,\, \tau_{P,|\mathcal{E}|},\, \nu_{P,|\mathcal{E}|}\bigr]\trans
    \in \mathbb{R}^{2P|\mathcal{E}| \times 1},
    \label{eq:theta_phys}
\end{equation}
where the entries corresponding to path $p$ of edge $e$ occupy indices $(e-1)\cdot 2P + 2p - 1$ for the delay and $(e-1)\cdot 2P + 2p$ for the Doppler in $\boldsymbol{\theta}_{\mathrm{phys}}$, for all edges.

\subsection{Joint Synchronization and Delay-Doppler Estimation via Linearized GaBP}
\label{sec:fine_tracking}

The key challenge in the joint estimation of the physical and synchronization parameters is that they are coupled in a nonlinear manner within the observation model, as shown in \eqref{eq:I/O_rel_final} and \eqref{eq:DISAC_DD_channel}.
However, it is possible to leverage the structure of the observation model to decouple the estimation of the two groups of parameters, $i.e.$, the physical delay-Doppler parameters and the synchronization parameters.
To solve the resulting bilinear estimation problem, we propose a \ac{GaBP} solver that can be applied after a first-order Taylor linearization of the observation model, which yields a linearized model with a separable structure that can be efficiently solved via message passing on a factor graph.

Since the observation depends on the unknowns only through the per-link effective indices $\ell_p^{\mathrm{eff}}$ and $f_p^{\mathrm{eff}}$, which by \eqref{eq:tau_eff}-\eqref{eq:nu_eff} decompose additively into a physical and a synchronization contribution, the physical parameters $\boldsymbol{\theta}_{\mathrm{phys}}$ \eqref{eq:theta_phys} are shared by both directions of an edge through \ac{TDD} reciprocity, whereas the synchronization parameters $\boldsymbol{\theta}_{\mathrm{sync}}$ \eqref{eq:theta_sync} enter anti-symmetrically across the two link directions. 
Thus, the observation is a known-gain nonlinear function of the stacked vector
\begin{equation}
    \boldsymbol{\theta} = [\boldsymbol{\theta}_{\mathrm{phys}}\trans,\; \boldsymbol{\theta}_{\mathrm{sync}}\trans]\trans,
\end{equation}
which we linearize in the following, via a current estimate $\hat{\boldsymbol{\theta}}^{(i)}$.

At iteration $i$, the reconstructed signal at node $n$ is given by
\begin{equation}
    \hat{\mathbf{y}}_n^{(i)} = \sum_{n'=1}^{N} \hat{\bar{\mathbf{H}}}_{n,n'}^{(i)}\, \mathbf{x}_{n'},
\end{equation}
where the estimated effective channel of the link from $n'$ to $n$, evaluated at the known path gains, is defined as
\begin{equation}
    \hat{\bar{\mathbf{H}}}_{n,n'}^{(i)} = \mathbf{F}_K \Big( \sum_{p=1}^{P} h_{p,e}\, \mathbf{V}^{\hat{f}_{p,e}^{\mathrm{eff},(i)}} \mathbf{\Psi}(\hat{\ell}_{p,e}^{\mathrm{eff},(i)}) \Big) \mathbf{F}_K\herm .
\end{equation}

Then, the complex residual at node $n$ can be mapped to the real domain, in order to be compatible with the real-valued Gaussian message passing, which yields
\begin{equation}
    \mathbf{z}_n^{(i)} \triangleq \begin{bmatrix} \mathrm{Re}(\mathbf{y}_n - \hat{\mathbf{y}}_n^{(i)}) \\ \mathrm{Im}(\mathbf{y}_n - \hat{\mathbf{y}}_n^{(i)}) \end{bmatrix} \in \mathbb{R}^{2K \times 1},
\end{equation}
and stacked over all nodes into the global residual given by
\begin{equation}
    \mathbf{z}^{(i)} \triangleq [\mathbf{z}_1^{(i)\trans},\, \cdots,\, \mathbf{z}_N^{(i)\trans}]\trans \in \mathbb{R}^{2NK \times 1}.
    \label{eq:global_residual}
\end{equation}

Finally, to linearize the residual, we perform a first-order Taylor expansion, which yields the separable bilinear model given by \vspace{-1ex}
\begin{equation}
    \mathbf{z}^{(i)} \approx \mathbf{H}_{\mathrm{phys}}\, \Delta\boldsymbol{\theta}_{\mathrm{phys}} + \mathbf{H}_{\mathrm{sync}}\, \Delta\boldsymbol{\theta}_{\mathrm{sync}} + \mathbf{w},
    \label{eq:linearized}
\end{equation}
with $\Delta\boldsymbol{\theta} = \boldsymbol{\theta} - \hat{\boldsymbol{\theta}}^{(i)}$ and $\mathbf{w} \sim \mathcal{N}(\mathbf{0}, \frac{N_0}{2}\mathbf{I}_{2NK})$. 

The Jacobians $\mathbf{H}_{\mathrm{phys}} \in \mathbb{R}^{2NK \times 2P|\mathcal{E}|}$ and $\mathbf{H}_{\mathrm{sync}} \in \mathbb{R}^{2NK \times 2(N-1)}$ isolate the sensitivity of the observation to the delay, Doppler and to the synchronization parameters, respectively, which allows the bivariate \ac{GaBP} solver to treat the two groups as coupled but distinct sets of variable nodes.
Both blocks are populated by iterating over the directed links $l \in \mathcal{L}$. 
For link $l$ with transmitter $n'_l$, receiver $n_l$ and edge $e_l$, the partial derivatives of $\hat{\mathbf{y}}_{n_l}$ with respect to $\tau_{p,e_l}$ and $\nu_{p,e_l}$ {\color{black} are} given by
\begin{subequations}
\begin{equation}
    \mathbf{j}_{\tau,p} = \frac{1}{T_s}\, \mathbf{F}_K \big( h_{p,e_l}\, \mathbf{V}^{\hat{f}_{p}^{\mathrm{eff}}} \dot{\mathbf{\Psi}}(\hat{\ell}_{p}^{\mathrm{eff}}) \big)\mathbf{F}_K\herm\, \mathbf{x}_{n'_l},
    \label{eq:j_tau}
\end{equation}
\begin{equation}
    \mathbf{j}_{\nu,p} = KT_s\, \mathbf{F}_K \big( h_{p,e_l} (-\jmath\frac{2\pi}{K}) \mathbf{D}\, \mathbf{V}^{\hat{f}_{p}^{\mathrm{eff}}} \mathbf{\Psi}(\hat{\ell}_{p}^{\mathrm{eff}}) \big)\mathbf{F}_K\herm\, \mathbf{x}_{n'_l},
    \label{eq:j_nu}
\end{equation}
\end{subequations}
with $\mathbf{D} \triangleq \mathrm{diag}([0,1,\ldots,K-1])$ and $\dot{\mathbf{\Psi}}$ the derivative of the delay matrix, which follows directly from the pulse derivative as \vspace{-1ex}
\begin{equation}
    [\dot{\mathbf{\Psi}}(\ell)]_{i,j} = -\dot{g}(i-j-\ell), 
    \label{eq:psi_dot_cd}
\end{equation}
where, for the \ac{RRC} pulse with roll-off $\alpha$, $\dot{g}(\cdot)\triangleq \frac{\mathrm{d}g(t)}{\mathrm{d}t}$ is available in closed form \cite{Proakis_2001}.

The corresponding per-link derivatives are assembled into the global Jacobians as follows. 
Both $\mathbf{H}_{\mathrm{phys}}$ and $\mathbf{H}_{\mathrm{sync}}$ are sparse and are initialized to zero, after which each link $l$ inserts its sensitivity blocks into the $2K$ rows of its receiver $n_l$ and the columns of the parameters it senses, such that these inserted blocks constitute the only nonzero entries of the two matrices. In particular, the real-valued $2K \times 2$ physical sensitivity block of path $p$ of link $l$ is given by
\begin{equation}
    \mathbf{J}_{\mathrm{phys},p}^{(l)} \triangleq \begin{bmatrix} \mathrm{Re}(\mathbf{j}_{\tau,p}) & \mathrm{Re}(\mathbf{j}_{\nu,p}) \\ \mathrm{Im}(\mathbf{j}_{\tau,p}) & \mathrm{Im}(\mathbf{j}_{\nu,p}) \end{bmatrix} \in \mathbb{R}^{2K \times 2},
    \label{eq:Hphys_populate}
\end{equation}
which is inserted into the rows of receiver $n_l$ and the two columns of $\boldsymbol{\theta}_{\mathrm{phys}}$ associated with path $p$ of edge $e_l$, $i.e.$, the indices $(e_l-1)2P+2p-1$ and $(e_l-1)2P+2p$ from \eqref{eq:theta_phys}. 

By \ac{TDD} reciprocity, the two directed links of edge $e_l$ fill these same columns within the rows of their respective receivers, such that both transmissions jointly inform the shared physical parameters.
Similarly, the synchronization sensitivities of link $l$ are aggregated over its paths as
\begin{subequations}
\begin{equation}
    \mathbf{J}_{\mathrm{rel},t}^{(l)} \triangleq \big[ \mathrm{Re}(\textstyle\sum_{p} \mathbf{j}_{\tau,p})\trans,\; \mathrm{Im}(\textstyle\sum_{p} \mathbf{j}_{\tau,p})\trans \big]\trans,
    \label{eq:Jrelt}
\end{equation}
\begin{equation}
    \mathbf{J}_{\mathrm{rel},f}^{(l)} \triangleq \big[ \mathrm{Re}(\textstyle\sum_{p} \mathbf{j}_{\nu,p})\trans,\; \mathrm{Im}(\textstyle\sum_{p} \mathbf{j}_{\nu,p})\trans \big]\trans,
    \label{eq:Jrelf}
\end{equation}
\end{subequations}
which, following \eqref{eq:tau_eff}-\eqref{eq:nu_eff}, enter with slope $+1$ at the transmitter and $-1$ at the receiver. 
Accordingly, within the rows of receiver $n_l$, the pair $[\mathbf{J}_{\mathrm{rel},t}^{(l)},\, \mathbf{J}_{\mathrm{rel},f}^{(l)}]$ is added to the columns $c_{n'_l}$ and $c_{n'_l}+1$ of the transmitter $n'_l$ and subtracted from the columns $c_{n_l}$ and $c_{n_l}+1$ of the receiver $n_l$, with the column indices $c_n$ as defined in \eqref{eq:theta_sync}, while the reference node $1$ has no associated columns. 
Since a node participates in several links, the contributions of all links incident to it accumulate in its columns, and repeating the insertion for every $l \in \mathcal{L}$ completes the Jacobians $\mathbf{H}_{\mathrm{phys}}$ and $\mathbf{H}_{\mathrm{sync}}$ of the linearized model \eqref{eq:linearized}.

Finally, because the delay and Doppler columns scale as $1/T_s$ and $KT_s$ and thus differ by many orders of magnitude, each Jacobian is scaled to unit column norm, to precondition the system, in order to avoid numerical issues in the \ac{GaBP} solver, which leads to
\begin{subequations}
    \begin{equation}
        \tilde{\mathbf{H}}_{\mathrm{phys}} = \mathbf{H}_{\mathrm{phys}} \mathbf{W}_{\mathrm{phys}}^{-1},
    \end{equation}
    \begin{equation}
        \tilde{\mathbf{H}}_{\mathrm{sync}} = \mathbf{H}_{\mathrm{sync}} \mathbf{W}_{\mathrm{sync}}^{-1},
    \end{equation}
\end{subequations}
with $\mathbf{W}_{(\cdot)} = \mathrm{diag}(w_{(\cdot),1},\ldots)$ and $w_{(\cdot),k} = \|\mathbf{H}_{(\cdot)}[:,k]\|_2 + \varepsilon$, while the prior variances of the parameters are scaled accordingly, given by
\begin{equation}
    \tilde{\phi}_{(\cdot),k} = \phi_{(\cdot),k}\, w_{(\cdot),k}^2.
    \label{eq:phi_scaled}
\end{equation}

After linearization and preconditioning, the problem is solved via a bivariate \ac{GaBP} that treats the physical and synchronization parameters as two distinct groups of variable nodes.
In the first soft-\ac{IC} step, the physical contribution is first cancelled, while the synchronization part is given by
\begin{equation}
    \mathbf{z}_{\mathrm{sync}} = \mathbf{z}^{(i)} - \tilde{\mathbf{H}}_{\mathrm{phys}}\, \tilde{\boldsymbol{\theta}}_{\mathrm{phys}},
    \label{eq:zsync}
\end{equation}
such that the soft-\ac{IC} residual targeting variable $l$ is defined as
\begin{equation}
    \tilde{z}_{m \to l} = [\mathbf{z}_{\mathrm{sync}}]_m - {\color{black}\sum_{j \neq l} [\tilde{\mathbf{H}}_{\mathrm{sync}}]_{m,j}\, \tilde{\theta}_{\mathrm{sync},j}},
    \label{eq:ztilde_sync}
\end{equation}
with the effective variance, given by
\begin{equation}
        v_{m \to l} \!=\! \frac{N_0}{2} \!+\! \sum_{k} [\tilde{\mathbf{H}}_{\mathrm{phys}}]_{m,k}^2\, \psi_{\mathrm{phys},k} \!+\! {\color{black}\sum_{j \neq l} [\tilde{\mathbf{H}}_{\mathrm{sync}}]_{m,j}^2\, \psi_{\mathrm{sync},j}}.
    \label{eq:vtilde_sync}
\end{equation}

Under the \ac{SGA}, the factor-to-variable precision and precision-weighted mean messages are defined as
\begin{subequations}
\begin{equation}
    \Lambda_{m \to l} = \frac{[\tilde{\mathbf{H}}_{\mathrm{sync}}]_{m,l}^2}{v_{m \to l}},
\end{equation}
\begin{equation}
    \mu_{m \to l} = \frac{[\tilde{\mathbf{H}}_{\mathrm{sync}}]_{m,l}\, \tilde{z}_{m \to l}}{v_{m \to l}},
    \label{eq:messages_sync}
\end{equation}
\end{subequations}
which are then used at the variable node to update the posterior mean and variance, where the posterior precision{\color{black} , which} aggregates all factor messages and the prior{\color{black} ,} is given by
\begin{equation}
    \Lambda_l = \frac{1}{\tilde{\phi}_{\mathrm{sync},l}} + \sum_{m=1}^{{\color{black} 2NK}} \Lambda_{m \to l},
    \label{eq:Lambda_sync}
\end{equation}
and the mean and variance are updated with damping $\rho$,
\begin{subequations}
\begin{equation}
    \tilde{\theta}_{\mathrm{sync},l} \leftarrow \rho\, \tilde{\theta}_{\mathrm{sync},l} + (1-\rho)\, \frac{1}{\Lambda_l} \sum_{m} \mu_{m \to l},
    \label{eq:theta_sync_update}
\end{equation}
\begin{equation}
    \psi_{\mathrm{sync},l} \leftarrow \rho\, \psi_{\mathrm{sync},l} + (1-\rho)\, \frac{1}{\Lambda_l}.
    \label{eq:psi_sync_update}
\end{equation}
\end{subequations}

For the physical parameter phase, the process is identical with the roles of the two blocks exchanged, such that the residual is given by \vspace{-1ex}
\begin{equation}
    \mathbf{z}_{\mathrm{phys}} = \mathbf{z}^{(i)} - \tilde{\mathbf{H}}_{\mathrm{sync}}\, \tilde{\boldsymbol{\theta}}_{\mathrm{sync}},
    \label{eq:zphys}
    \vspace{-1ex}
\end{equation}
and its effective variance retains the cross-domain synchronization uncertainty,
\begin{equation}
    v_{m \to k} \!=\! \frac{N_0}{2} \!+\! {\color{black}\sum_{s} [\tilde{\mathbf{H}}_{\mathrm{sync}}]_{m,s}^2\, \psi_{\mathrm{sync},s}} \!+\! {\color{black}\sum_{j \neq k} [\tilde{\mathbf{H}}_{\mathrm{phys}}]_{m,j}^2\, \psi_{\mathrm{phys},j}}.
    \label{eq:v_phys}
\end{equation}

Next, after the outer loops, the cross-domain terms left in \eqref{eq:vtilde_sync} inflate the posterior variances, such that a final dedicated pass is required that treats each converged group as a hard decision and removes the cross-domain term. 
For the physical variables, the synchronization estimate is hard-subtracted, which leads to \vspace{-1ex}
\begin{equation}
    \mathbf{z}'_{\mathrm{phys}} = \mathbf{z}^{(i)} - \tilde{\mathbf{H}}_{\mathrm{sync}}\, \tilde{\boldsymbol{\theta}}_{\mathrm{sync}},
    \label{eq:z_prime_phys}
\end{equation}
with noise floor reduced to
\begin{equation}
    v_{m \to k}^{\mathrm{pol}} = \frac{N_0}{2} + {\color{black}\sum_{j \neq k} [\tilde{\mathbf{H}}_{\mathrm{phys}}]_{m,j}^2\, \psi_{\mathrm{phys},j}}.
    \label{eq:v_polish_phys}
    \vspace{-1ex}
\end{equation}

Similarly, for the synchronization variables, the physical contribution is hard-subtracted, which yields
\begin{equation}
    \mathbf{z}'_{\mathrm{sync}} = \mathbf{z}^{(i)} - \tilde{\mathbf{H}}_{\mathrm{phys}}\, \tilde{\boldsymbol{\theta}}_{\mathrm{phys}},
    \label{eq:z_prime_sync}
\end{equation}
\begin{equation}
    v_{m \to l}^{\mathrm{pol}} = \frac{N_0}{2} + {\color{black}\sum_{j \neq l} [\tilde{\mathbf{H}}_{\mathrm{sync}}]_{m,j}^2\, \psi_{\mathrm{sync},j}}.
    \label{eq:v_polish_sync}
\end{equation}

Finally, the physical-unit corrections can be retrieved by un-scaling the preconditioned estimates, given by
\begin{subequations}
\begin{equation}
    \Delta\hat{\epsilon}_{t,n} = \frac{[\Delta\tilde{\boldsymbol{\theta}}_{\mathrm{sync}}]_{2(n-2)+1}}{w_{\mathrm{sync},\, 2(n-2)+1}}, \quad \Delta\hat{\epsilon}_{f,n} = \frac{[\Delta\tilde{\boldsymbol{\theta}}_{\mathrm{sync}}]_{2(n-2)+2}}{w_{\mathrm{sync},\, 2(n-2)+2}},
    \label{eq:unscale_sync}
\end{equation}
\begin{equation}
    \Delta\hat{\tau}_{p,e}\!\! = \!\!\frac{[\Delta\tilde{\boldsymbol{\theta}}_{\mathrm{phys}}]_{(e-1)2P+2p-1}}{w_{\mathrm{phys}, (e-1)2P+2p-1}}, \!\!\!\!\quad \Delta\hat{\nu}_{p,e} \!\!=\!\!\frac{[\Delta\tilde{\boldsymbol{\theta}}_{\mathrm{phys}}]_{(e-1)2P+2p}}{w_{\mathrm{phys}, (e-1)2P+2p}},
    \label{eq:unscale_phys}
\end{equation}
\end{subequations}
which can be used to update the expansion point as $\hat{\boldsymbol{\theta}}^{(i+1)} = \hat{\boldsymbol{\theta}}^{(i)} + \Delta\hat{\boldsymbol{\theta}}$, and yield the final estimate $\hat{\boldsymbol{\Theta}}^{(I_{\mathrm{macro}})}$ after convergence.

\vspace{-0.5ex}
\section{Numerical Results}
\label{sec:results}
\vspace{-0.5ex}

We evaluate the proposed framework over a \ac{DISAC} network of $N=5$ single-antenna nodes with all $N(N-1)=20$ directed links active over shared time-frequency resources, using {\color{black}$K=2048$} \ac{QPSK} pilots per block, $P=2$ paths per link, $f_c=3\,\mathrm{GHz}$ and {\color{black}$B=20\,\mathrm{MHz}$}. 
While node $1$ is chosen as the reference clock, the non-reference offsets are drawn as $\epsilon_{t,n}\sim\mathcal{U}(\pm0.1\,T_s)$ and $\epsilon_{f,n}\sim\mathcal{U}(\pm20\,\mathrm{Hz})$, while the per-path delays and Dopplers follow from the geometric model with $d_p\sim\mathcal{U}(100,1600)\,\mathrm{m}$, $v_p\sim\mathcal{U}(\pm30)\,\mathrm{m/s}$. 
%
%
For the sake of comparison, we calculate the normalized \ac{RMSE} of the synchronization parameters $\epsilon_t,\epsilon_f$, and the radar parameters $\tau,\nu$, over a \ac{SNR} across $-10$ to $30\,\mathrm{dB}$.

\begin{figure}[H]
    \centering
    \includegraphics[width=\columnwidth]{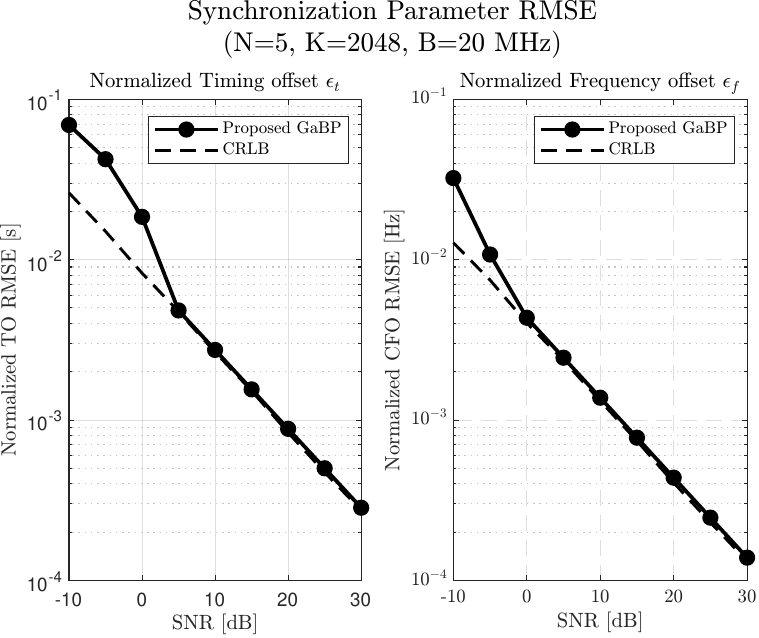}
    \caption{{\color{black} Normalized RMSE of TO and CFO vs.\ SNR}}
    \label{fig:Sync_estimation_error}
\end{figure}

\vspace{-2ex}
Since, to the best of our knowledge, no existing scheme jointly estimates the network synchronization offsets together with the delay-Doppler parameters in a distributed \ac{ISAC} setting, we benchmark the proposed estimator against the \ac{CRLB} rather than against a competing method.
Under the given assumptions, the joint \ac{FIM} is defined as\vspace{-1ex}
\begin{equation}
\color{black}
    \text{FIM} = \frac{2}{N_0}\, \big[\mathbf{H}_{\mathrm{phys}}\, \mathbf{H}_{\mathrm{sync}}\big]\trans\big[\mathbf{H}_{\mathrm{phys}}\, \mathbf{H}_{\mathrm{sync}}\big].
    \label{eq:crlb_fim}
\vspace{-2ex}
\end{equation}

Figure \ref{fig:Sync_estimation_error} illustrates the {\color{black} estimation performance for the} synchronization parameters, $i.e.$, \ac{TO} and \ac{CFO}.
It can be observed that the proposed method works well for high \ac{SNR}, closely matching the \ac{CRLB}, while it slightly deviates from it for lower \ac{SNR}.
More specifically, for the \ac{TO}, the error {\color{black} deviates from the \ac{CRLB} below $5\,\mathrm{dB}$}, while the \ac{CFO} estimate only deviates {\color{black} below $0\,\mathrm{dB}$}.

\begin{figure}[H]
    \centering
    \includegraphics[width=\columnwidth]{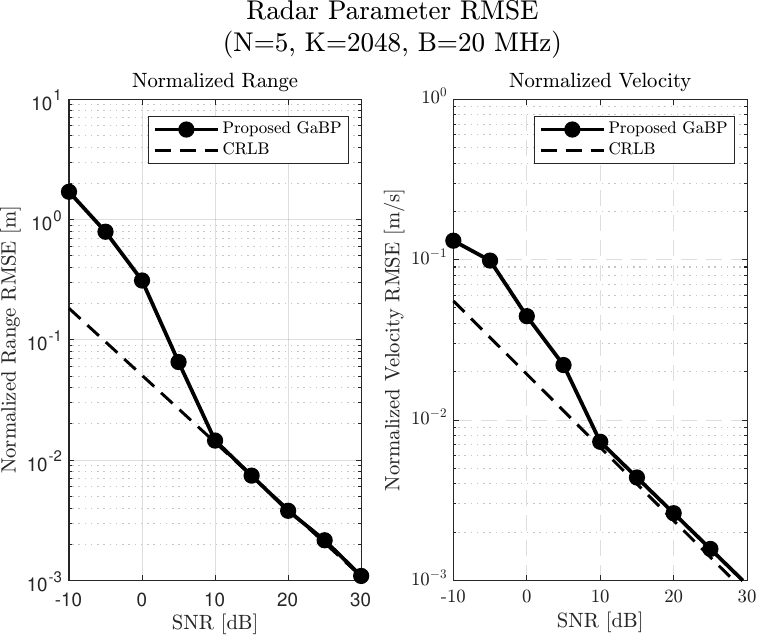}
    \caption{{\color{black} Normalized RMSE of range and velocity vs.\ SNR}}
    \label{fig:Radar_estimation_error}
\end{figure}

Next, Figure \ref{fig:Radar_estimation_error} demonstrates the radar parameter estimation performance.
While the proposed algorithm yields the delay and Doppler parameters, for comparison, we converted them to range and velocity {\color{black} according to \eqref{eq:tau_from_d} and \eqref{eq:nu_from_v}}.
It can be observed that both range and velocity estimates match the \ac{CRLB} for high \ac{SNR}, followed by a short waterfall behavior, until they stabilize again at $10\,\mathrm{dB}$, for both estimates.

\vspace{-1ex}
\section{Conclusion}
\label{sec:conclusion}
\vspace{-1ex}
{\color{black} We proposed a joint synchronization and radar parameter estimation framework for distributed \ac{OFDM}-\ac{ISAC} networks operating in \ac{DD} channels, which exploits the structural coupling between the two problems via a linearized observation model and a bivariate \ac{GaBP} solver that iteratively refines \ac{TO}, \ac{CFO}, delay, and Doppler estimates jointly. 
Simulation results confirm that the proposed estimator closely approaches the \ac{CRLB} under known path gains and a coarse initialization, while future work will incorporate a preceding coarse-estimation stage that provides the initial delay-Doppler estimates and the path gains, and will extend the evaluation across varying network topologies and waveforms.}

    \vspace{-2ex}

\end{document}